\documentclass[aps,prb,longbibliography,superscriptaddress,preprint]{revtex4-2}
\usepackage{mathrsfs}
\usepackage{amsmath}
\usepackage{amsfonts}
\usepackage{amssymb}
\usepackage{amsthm}
\usepackage{graphicx}
\usepackage{color}
\usepackage{hyperref}
\usepackage{bm}
\usepackage[caption=false]{subfig}
\usepackage{verbatim}
\usepackage{ulem}

\begin{document}
\title{Interfacial control of vortex-limited critical current in type II superconductor films}
\author{Marius K. Hope}
\affiliation{Center for Quantum Spintronics, Department of Physics, Norwegian University of Science and Technology, NO-7491 Trondheim, Norway}
\author{Morten Amundsen}
\affiliation{Nordita, KTH Royal Institute of Technology and Stockholm University, Hannes Alfv{\'e}ns v{\"a}g 12, SE-106 91 Stockholm, Sweden}

\author{Dhavala Suri}
\affiliation{Francis Bitter Magnet Laboratory \& Plasma Science and Fusion Center, Massachusetts Institute of Technology, Cambridge, MA-02139, USA}
\author{Jagadeesh S. Moodera}
\affiliation{Francis Bitter Magnet Laboratory \& Plasma Science and Fusion Center, Massachusetts Institute of Technology, Cambridge, MA-02139, USA}

\affiliation{Department of Physics, Massachusetts Institute of Technology, Cambridge, MA-02139,USA}
\author{Akashdeep Kamra}
\email{akashdeep.kamra@uam.es}
\affiliation{Condensed Matter Physics Center (IFIMAC) and Departamento de F\'{i}sica Te\'{o}rica de la Materia Condensada, Universidad Aut\'{o}noma de Madrid, E-28049 Madrid, Spain} 
\affiliation{Center for Quantum Spintronics, Department of Physics, Norwegian University of Science and Technology, NO-7491 Trondheim, Norway}

\begin{abstract}
In a small subset of type II superconductor films, the critical current is determined by a weakened Bean-Livingston barrier posed by the film surfaces to vortex penetration into the sample. A {film} property thus depends sensitively on the surface or interface to an adjacent material. We theoretically investigate the dependence of vortex barrier and critical current in such films on the Rashba spin-orbit coupling at their interfaces with adjacent materials. Considering an interface with a magnetic insulator, we find the spontaneous supercurrent resulting from the {exchange} field and interfacial spin-orbit coupling to substantially modify the vortex surface barrier, {consistent with a previous prediction}. Thus, we show that the critical currents in superconductor-magnet heterostructures can be controlled, and even enhanced, via the interfacial spin-orbit coupling. Since the latter can be controlled via a gate voltage, our analysis predicts a class of heterostructures amenable to gate-voltage modulation of superconducting critical currents. It also sheds light on the recently observed gate-voltage enhancement of critical current in NbN superconducting films.
\end{abstract}


\maketitle

\section{Introduction}\label{sec:intro}

The superconducting phase emerges when pairs of conduction electrons experiencing an attraction condense into a macroscopic quantum coherent state~\cite{deGennes1966,Parks2019}. Below a critical temperature, often abbreviated to Tc, the superconducting phase is lower in energy than the normal metallic state. This macroscopic coherence supports dissipationless flow of charge up to a maximum critical current density above which superconductivity is destroyed and the system reverts to its normal metal state~\cite{deGennes1966,Parks2019}. The virtually absent dissipation and macroscopic coherence underlie the central role superconductors are playing in various emerging quantum technologies~\cite{Xiang2013,Krantz2019}. The major fraction of these employ the so-called conventional superconductors that emerge from metallic states with very high charge carrier densities and are well-understood within the Bardeen-Cooper-Schrieffer (BCS) theory~\cite{Bardeen1957}.

In contrast, a large variety of superconductors, broadly called ``unconventional'', emerge from insulating or low electron density normal states~\cite{Bennemann2008}. The physics of these varies from one system to another and only a few similarities can be noted. Among them is their demonstrated tunability via an applied gate voltage or an electric field~\cite{Ahn2003,Caviglia2008,Frey1995,Xi1992,Ueno2014}, which can significantly modify the carrier concentration and thus, various superconducting properties such as Tc. The low carrier density is of fundamental importance in this gate voltage control for two reasons. First, an electric field generated by an adjacent gate electrode can only exist in a semiconducting or insulating system. In typical metals, an applied electric field is screened out by the charge carriers on atomic length scales~\cite{Bonfiglioli1956,Bonfiglioli1959,Lipavsky2010,Piatti2018}. Second, the relative change in carrier concentration induced by an applied gate voltage is negligible for metallic systems due to their very high pre-existing carrier densities~\cite{Glover1960,Bonfiglioli1956}.

Hence, it came as a (pleasant) surprise when the critical current in a range of metallic conventional superconductor films responded significantly to gate voltages~\cite{Simoni2018,Paolucci2018,Puglia2020,Rocci2020,Ritter2021,Rocci2021,Puglia2021,Alegria2021}. Two key features should be noted. First, the modification observed in critical current is even in the gate voltage, i.e., it depends only on the electric field strength. Second, Tc remained unchanged with the gate voltage. Several potential mechanisms such as quasiparticle injection~\cite{Ritter2021,Alegria2021}, disorder~\cite{Paolucci2018}, Sauter-Schwinger effect~\cite{Solinas2021}, and others~\cite{Virtanen2019,Mercaldo2020,Ritter_arxiv} have been considered to explain the observed suppression of the critical current with the gate voltage. However, explanations consistent with all the key experimental features and a broad consensus for each material system are still being pursued~\cite{Puglia2021}. Furthermore, the recently observed {\it enhancement} in the critical current of type II superconductor NbN films~\cite{Rocci2021} stands out. While it shares the even-in-gate-voltage dependence and no-Tc-change feature with other observations, none of the mechanisms speculated above can justify an increase in the critical current, which has been attributed to a vortex surface barrier based mechanism~\cite{Rocci2021,Shmidt1970,Shmidt1970b}. A systematic theory of these observations should aim to explain the specific gate voltage dependencies of Tc and critical current. At the same time, it should clarify what system property (e.g., carrier density or interfacial electric field) is primarily influenced by the gate voltage and how this change affects the superconductor properties~\cite{Piatti2021}.

In this {article, taking inspiration from the above mentioned experiments, we theoretically investigate the vortex mechanism of critical current control, put forth as a possible origin of the observed enhancement in NbN films~\cite{Rocci2021}. In this process, we identify the key physics at play, finding it to be consistent with the main experimental features mentioned above. However, our focus is on investigating} superconductor$|$magnetic insulator (S$|$MI) hybrids {because they are found to} follow similar physics, {be} simpler to analyze theoretically, and manifest a potentially stronger gate-voltage control or enhancement of superconducting critical currents. {Our choice of investigating such hybrids is additionally inspired by a recent work due to Mironov and Buzdin, which found superconductor$|$ferromagnetic insulator interfaces to host spontaneous supercurrents~\cite{Mironov2017}. They further commented on a consequent renormalization of the vortex surface barrier by the induced spontaneous supercurrent. Here, following similar considerations~\cite{Mironov2017,Olthof2019}, we theoretically demonstrate a control of vortex surface barriers in a range of S$|$MI hybrids, where MI can be a ferro- or an antiferromagnet~\cite{Kamra2018}. More significantly, we delineate the critical current dependence on these vortex barriers generalizing the work of Shmidt~\cite{Shmidt1970,Shmidt1970b} in several crucial aspects. Such S$|$MI hybrids} have already become the workhorse for superconducting spintronics (e.g., see Refs.~\cite{Bergeret2018,Eschrig2015,Linder2015,Yao2018,Hijano2021,Rouco2019,Heikkila2019,Diesch2018}) and constitute a mature conveniently-fabricated experimental platform. We exclusively consider metallic conventional superconductors that are well described via BCS and Ginzburg-Landau (GL) theories~\cite{deGennes1966,Parks2019}. Furthermore, {as we focus on the vortex mechanism of critical current~\cite{Shmidt1970,Shmidt1970b}, we} consider type II superconductors.

Our first key premise, which could have a broader relevance for the gate voltage control and experiments mentioned above, is that the gate voltage primarily influences the interfacial Rashba spin-orbit coupling (SOC)~\cite{Nitta1997,Manchon2015}. It induces a strong electric field that is screened within atomic length scales from the superconductor surface~\cite{Piatti2021,Lipavsky2010}. However, the finite voltage drop across the infinitesimal interface causes a significant gate voltage-induced interfacial Rashba SOC~\cite{Liu2014}. This assumption is supported by the recent observation of gate voltage-controlled interfacial spin-orbit torques in a metallic ferromagnet/semiconductor heterostructure~\cite{Chen2018}. The gate voltage was found to induce a strong Rashba SOC at such an interface. Furthermore, gate voltage modulation of Rashba SOC in superconducting phases living on oxide interfaces has previously been shown to underlie superconductivity modulation~\cite{Caviglia2008,Caviglia2010,Ben2010}. Our second key premise is that the critical current in our type II superconductor films is the value at which the Lorentz force on vortices spontaneously nucleated at one side overcomes the Bean-Livingston surface barrier~\cite{Bean1964}. At this current, these vortices are able to traverse the film thereby causing {dissipation and destroying superconductivity~\cite{Shmidt1970,Shmidt1970b}, in the sense that a nonzero voltage develops. Further, the dissipation causes heating and may subsequently kill the superconductivity thermally.} Thus, {the critical current} is determined and controlled directly by the surfaces. This assumption is also supported by several theoretical and experimental works consistent with this vortex mechanism of critical current in {some} type II superconducting films~\cite{Mawatari1994,Aslamazov1975,Clem2011,Ilin2014,Stejic1994,Lara2017}. {However, this is only one of the mechanisms that can determine critical current in a film, as discussed further below.}

Thus, employing the GL phenomenology, we analytically evaluate the vortex surface barrier and the ensuing critical current in type II superconductor films interfaced with magnetic insulators. In the presence of interfacial Rashba SOC and effective {exchange} field induced by the adjacent magnet, a spontaneous supercurrent flows at the interface, as shown recently by Mironov and Buzdin~\cite{Mironov2017}. It substantially modifies the vortex surface barrier thereby increasing or decreasing the critical current, depending on the relative orientations of induced {exchange} field, applied magnetic field, and transport current through the film. The modification in vortex barrier and critical current is proportional to the Rashba SOC strength, which can be tuned by a gate voltage. The critical temperature Tc of the superconducting state remains unaltered as per this mechanism. For superconductor films without adjacent magnets~\cite{Rocci2021}, the critical current change is second order in the Rashba SOC. It is thus expected to be weaker than in superconductor-magnet hybrids and even in the applied gate voltage.

\section{Overview and Model}\label{sec:overview}

In this section, we provide a qualitative discussion of the key phenomena at play and introduce the general mathematical framework for the problem. Our goal here is to provide an intuitive understanding and an overview of the methodology rather than the full details.

Different mechanisms can determine the critical current in superconducting films~\cite{Shmidt1970b,Bardeen1962,Likharev1979,deGennes1966}. The most well-known is the critical depairing mechanism, which causes a rapid destruction of the Cooper pairs when their kinetic energy begins to exceed the superconducting condensation energy~\cite{Bardeen1962,deGennes1966}. Other mechanisms, such as weakest-link~\cite{Likharev1979}, begin to dominate films with some disorder and granular structure. In type II superconductor films with thicknesses larger than the coherence length, there is an additional mechanism which can  become dominant in clean films~\cite{Shmidt1970,Shmidt1970b}. Vortices~\cite{Abrikosov1957,Abrikosov1964} have a tendency to be nucleated and annihilated at the surfaces of a type II superconductor film. These are unable to penetrate the film due to the so-called Bean-Livingston surface barrier~\cite{Bean1964,deGennes1966,Brandt2013}. In the presence of a transport current through the film, these vortices additionally experience a Lorentz force which tends to pull them from one side to the other. At large enough current, this Lorentz force exceeds its counterpart due to the Bean-Livingston barrier. Consequently, vortices nucleated at one surface move across the superconducting film to the other side causing dissipation and loss of superconductivity~\cite{Shmidt1970}. The dominance of this mechanism requires two main properties - a weak Bean-Livingston surface barrier and an absence of vortex (or flux) pinning centers in the superconducting film. This situation is likely to happen when the interface is somewhat disordered, thereby lowering the barrier, and the ``bulk'' is clean, thereby preventing disorder that could pin the vortices. In this work, we consider such films and assume the vortex instability to determine the critical current. {Further discussion on which films may be dominated by this mechanism is provided in Sec.~\ref{sec:discussion}.}

\begin{figure}[tb]
	\centering
	\includegraphics[width=85mm]{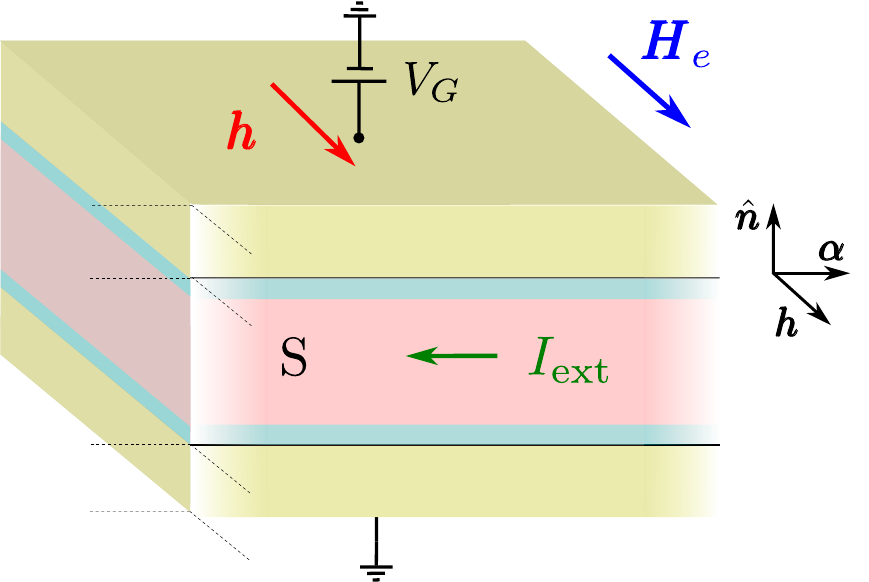}
	\caption{Schematic depiction of the system under consideration. A type II superconductor film S (pink) is sandwiched between two insulators (yellow). An in-plane magnetic field $\pmb{H}_e$ is applied and a current $I_{\mathrm{ext}}$ is driven through the film. Interfacial Rashba spin-orbit coupling (SOC) is assumed to exist in thin layers (blue) adjoining the interfaces. Its strength can be tuned via an applied gate voltage $V_G$. When the adjacent insulators are magnetic, {exchange} field $\pmb{h}$ is induced in the interfacial layer (blue) of S. The critical current in such films depends on Rashba SOC and the induced {exchange} field. These can be controlled via the gate voltage $V_G$ and the applied magnetic field $\pmb{H}_e$ providing multiple handles on critical current in the superconductor. {The coordinate system on the right depicts the orthogonal relation between the symmetry-breaking vector $\hat{\pmb{n}}$ on the upper S surface, the {exchange} field $\pmb{h}$, and the effective field $\pmb{\alpha}$ experienced by the S [Eq.~\eqref{eq:GL_free_energy}].}}
	\label{fig:Schematic_main}
\end{figure}

We employ the GL phenomenology to describe our superconductor film with thickness $d$, coherence length $\xi$, and London penetration depth $\lambda$, under the assumption $\xi \ll d \ll \lambda$. As depicted in Fig.~\ref{fig:Schematic_main}, an externally injected transport current $I_{\mathrm{ext}}$ flows parallel to the surfaces. An external magnetic field $\pmb{H}_e$ is applied tangential to the superconductor surfaces. The adjacent magnets and interfaces to them are incorporated in our model by including Rashba SOC and proximity-induced {exchange} field in a thin layer with thickness $l_{so}$ adjoining the interface (Fig.~\ref{fig:Schematic_main}). 

The GL free energy density $f(\pmb{r})$ at a position $\pmb{r}$ within the superconductor film is~\cite{Mironov2017,Samokhin2004,Kaur2005,Olthof2019,Dimitrova2007}:  
\begin{align}\label{eq:GL_free_energy}
f(\pmb{r}) & = a|\psi|^2 + \frac{b}{2}|\psi|^4 + \frac{\left| \hat{\pmb{D}} \psi \right|^2}{2m} + \frac{\pmb{B}^2}{2\mu_0}  +\frac{\pmb{\alpha}}{2m} \cdot \left( \psi^* \left( \hat{\pmb{D}} \psi \right) + \mathrm{h.c.} \right),
\end{align}
where $\psi$ is the superconducting order parameter, $a$ and $b$ are the standard GL coefficients describing a homogeneous superconductor, $\hat{\pmb{D}} \equiv -\textit{i}\hbar\nabla-2e\pmb{A}$ is the gauge-invariant momentum operator ($e<0$), $m$ is the effective mass of a Cooper pair, $\pmb{B} = \pmb{B}(\pmb{r}) = \nabla\times\pmb{A}$ is the spatially resolved magnetic flux density with $\pmb{A}$ the magnetic vector potential, and $\mu_0$ is the vacuum permeability. 

The last term in Eq.~\eqref{eq:GL_free_energy} above captures the free energy contribution due to the combined effect of Rashba SOC and a proximity-induced {exchange} field $\pmb{h}$~\cite{Samokhin2004,Dimitrova2007,Kaur2005}, and is parameterized via $\pmb{\alpha} = \pmb{\alpha}(\pmb{r}) = \beta(\pmb{r}) \left( \hat{\pmb{n}} \times \pmb{h} \right)$~\cite{Mironov2017,Olthof2019}. Here, $\beta(\pmb{r})$ is proportional to the Rashba SOC parameter $\alpha_{\mathrm{Ra}}$ that may depend on the position, and $\hat{\pmb{n}}$ is the unit vector along the direction of the spatial symmetry-breaking that causes Rashba SOC. $\beta$ further depends on the properties of the superconductor and will be detailed in the context of estimating effects using experimentally obtained parameters (Sec.~\ref{sec:estimation}). In our consideration of magnet-superconductor interfaces, we assume $\alpha_{\mathrm{Ra}}$ to be a nonzero constant in a thin layer with thickness $l_{so}$ next to the film surface (Fig.~\ref{fig:Schematic_main}), and $\hat{\pmb{n}}$ is normal to the interface directed outwards from the superconductor. This free energy contribution can be motivated as follows. Due to Rashba SOC, electrons with a given spin have lower energy if they move along a preferred direction. Thus, in the presence of an {exchange} field, which lowers the energy of electrons with a certain spin, the orbital motion of the Cooper pairs also develops a preference resulting in spontaneous supercurrents~\cite{Mironov2017}. The field $\pmb{h}$ needs to come from an {exchange}-induced spin-splitting via proximity to a magnet, and not an applied magnetic field. 

We work here in the London limit, i.e., the superconducting order parameter is assumed to have a constant magnitude: $\psi(\pmb{r}) \equiv \psi_0 e^{i\varphi(\pmb{r})}$. In this limit, our spatial resolution is assumed to be much larger than $\xi$ and vortices are treated as point objects via delta functions. Using Eq.~\eqref{eq:GL_free_energy}, the total free energy $F$ is thus expressed as~\cite{Olthof2019}
\begin{align}\label{eq:free_energy_vorticity}
F & =  \int d^3 r \left[ f_0 + \frac{1}{2\mu_0\lambda^2}\left\{\left(\pmb{\Phi}-\pmb{A}\right)^2 + \frac{\pmb{\alpha}}{e}\cdot\left(\pmb{\Phi}-\pmb{A}\right) + \lambda^2\pmb{B}^2\right\} \right],
\end{align}
where the integral runs over the superconductor volume $V_S$, $f_0 \equiv  a\psi_0^2  +  b \psi_0^4/2$ is a constant, $\pmb{\Phi}\equiv (\hbar/2e)\pmb{\nabla}\varphi$ is the vorticity~\cite{deGennes1966}, and $\lambda \equiv \sqrt{m/(4e^2\mu_0\psi_0^2)}$ is the London penetration depth. 

The equation governing the superconducting state is obtained by minimizing the total free energy [Eq.~\eqref{eq:free_energy_vorticity}] with respect to $\pmb{A}$~\footnote{In the considered London limit, minimization of the free energy with respect to $\psi$ becomes superfluous.}, which yields:
\begin{align}\label{eq:jgen}
\mu_0\lambda^2 \pmb{j} & = -\pmb{A} +\pmb{\Phi} + \frac{\pmb{\alpha}}{2e},
\end{align}
where we have additionally employed the Maxwell's equation for current density $\mu_0 \pmb{j} = \nabla \times \pmb{B}$.  Considering vortices with flux along $\hat{\pmb{z}}$ at positions $\pmb{\rho}_v$ in the x-y plane, we assume translational invariance along the z direction. With these assumptions, taking curl on both sides of Eq.~\eqref{eq:jgen} and employing Maxwell's equations, we obtain:
\begin{align}\label{eq:bmain}
\lambda^2 \nabla^2\pmb{B} - \pmb{B} & = -\phi_0 \hat{\pmb{z}} ~ \sum_{v} \delta_2(\pmb{\rho}-\pmb{\rho}_v) - \frac{\nabla\times\pmb{\alpha}}{2e},
\end{align}
where $\phi_0 \equiv h/(2e)$ is the magnetic flux quantum associated with each vortex and $\delta_2(\pmb{\rho})$ is the two-dimensional Dirac delta function. Employing Eq.~\eqref{eq:bmain} together with the boundary conditions resulting from the external magnetic field and the injected transport current, we can determine the magnetic flux density and supercurrent distribution in the superconductor for any assumed vortex configuration. These can then be substituted in Eq.~\eqref{eq:free_energy_vorticity} to obtain the Gibbs free energy density $G/V_S$ in a constant applied field $\pmb{H}_e$ averaged over the superconductor:
\begin{align}\label{eq:gibbs1}
\frac{G}{V_S} & =  \frac{1}{V_S} \int d^3 r \left[ \frac{1}{2\mu_0} \left[ \pmb{B}^2 + \lambda^2 (\pmb{\nabla} \times \pmb{B})^2-\frac{\pmb{\alpha}^2}{4e^2\lambda^2} \right]- \pmb{B} \cdot \pmb{H}_e \right],
\end{align}
where we have employed Eq.~\eqref{eq:bmain} and dropped the constant contribution $f_0$ in obtaining the expression above.

We are now ready to summarize our methodology. Assuming an applied magnetic field, externally injected transport supercurrent, and vortices located at positions $\pmb{\rho}_v$ in the superconductor, we evaluate the magnetic flux density $\pmb{B}(\pmb{r})$ by solving Eq.~\eqref{eq:bmain}. This solution is substituted in Eq.~\eqref{eq:gibbs1} to obtain the superconductor Gibbs free energy density as a function of the vortex positions $\pmb{\rho}_v$. The force per unit volume $\pmb{F}_s$ experienced by a vortex is given by the negative gradient of the Gibbs free energy density with respect to the vortex position $\pmb{\rho}_v$~\footnote{Shmidt accounted for the Lorentz force on the vortices by incorporating an effective term in the Gibbs free energy density~\cite{Shmidt1970}. Such an incorporation does not seem possible in the general case of arbitrary vortex locations. Thus, we treat Lorentz force in its direct form and separately from the force evaluated via the Gibbs free energy density gradient.}. To this, we need to add the Lorentz force per unit volume exerted by the externally injected transport current density $\pmb{j}_{\mathrm{ext}}$ making the net force per unit volume on a vortex:
\begin{align}\label{eq:fnet}
\pmb{F}_{\mathrm{net}} & = \pmb{F}_{s} + \frac{\phi_0 \pmb{j}_{\mathrm{ext}}(\pmb{\rho}_v) \times \hat{\pmb{z}}}{\mathcal{A}},
\end{align}   
where $\mathcal{A}$ is the area per vortex. Up to the critical current, $\pmb{F}_{\mathrm{net}}$ evaluated at the surface is directed outwards from the superconductor and vortices are not allowed to enter. Just above the critical current, $\pmb{F}_{\mathrm{net}}$ drives vortices nucleated at one surface to the other resulting in dissipation and loss of superconductivity. 

Thus, we formulate the following generalization of Shmidt's criterion~\cite{Shmidt1970} to define the critical current: 
\begin{align}\label{eq:criterion}
I_{c} & = \mathrm{min}_{\mathrm{vortex~axes}} \left[ \left. \pmb{F}_{\mathrm{net}}(I_{\mathrm{ext}})\right|_{\mathrm{surface}} \cdot \hat{\pmb{n}}  = 0 \right],
\end{align}
where the minimum value over all possible configurations for the vortex axes has to be chosen. The requirement of minimizing over all possible vortex axes arises when there is no preferred direction set by, for example, a large external applied field and is crucial in the above general criterion. For ease of calculations, we find it convenient to choose our coordinate frame with the vortex axis along $\hat{\pmb{z}}$ [Eq.~\eqref{eq:bmain}]. Nevertheless, our analysis puts no restrictions on the vortex axis direction in general. 

Following the analysis as outlined above and described in detail for certain cases of interest below, we obtain the following expression for the critical current in the superconducting film valid for a wide range of parameters:
\begin{align}
I_{c} = & \frac{2 W d^2}{4 \lambda^2} \mathrm{min} \left\{ \left(H_{sL} - H_e\right) , \left(H_{sR} + H_e\right)  \right\},  
\end{align}
where $W$ is the superconductor film width perpendicular to the direction of external current injection, and $H_{sL,sR}$ represent the vortex surface barrier fields on the left and right surfaces. The physics behind this expression will be clarified in the next sections. For now, we note that this is one of our main results which shows that the critical current in the superconductor can be influenced by controlling the vortex surface barrier fields $H_{sL,sR}$ via various interfacial effects.

With the methodology outlined above, we describe a superconductor film without any SOC in Sec.~\ref{sec:SwoSOC}. This corresponds to assuming $\pmb{\alpha} = \pmb{0}$ and provides a preliminary understanding of the critical current for the vortex mechanism considered here. In Sec.~\ref{sec:SwSOCZ}, we analyze a superconductor bearing Rashba SOC and {exchange} field resulting in a finite and general $\pmb{\alpha}(\pmb{r})$. The next section \ref{sec:S-MI} evaluates critical current in different superconductor-magnet hybrids. Here, interfaces with ferro- or antiferromagnets are assumed to provide {exchange} field and interfacial Rashba SOC giving rise to finite $\pmb{\alpha}$ in a thin layer next to the interface. As a result, the critical current is found to depend on the SOC strength, controllable via a gate voltage, and the {exchange} field, tuned via magnetic order in the adjacent materials. A robust control of critical current is thus predicted. In Sec.~\ref{sec:SwSOC}, we consider a superconductor film interfaced with nonmagnetic insulators such that it bears a gate voltage-tunable Rashba SOC but no {exchange} field. We argue from symmetry that the Rashba SOC alone gives a correction to critical current that scales as $\alpha_{\mathrm{Ra}}^2$. This dependence is compared with the experiments observing critical current enhancement in such superconducting films~\cite{Rocci2021}. We estimate the degree of critical current modulation using experimentally measured parameters in Sec.~\ref{sec:estimation}. Some of the assumptions and weaknesses of our theoretical model are discussed in Sec.~\ref{sec:discussion}, where comments on future efforts are also made. We conclude by summarizing our work in Sec.~\ref{sec:summary}.

\section{Superconductor film without SOC}\label{sec:SwoSOC}

\begin{figure}[tbh]
	\centering
	\includegraphics[width=160mm]{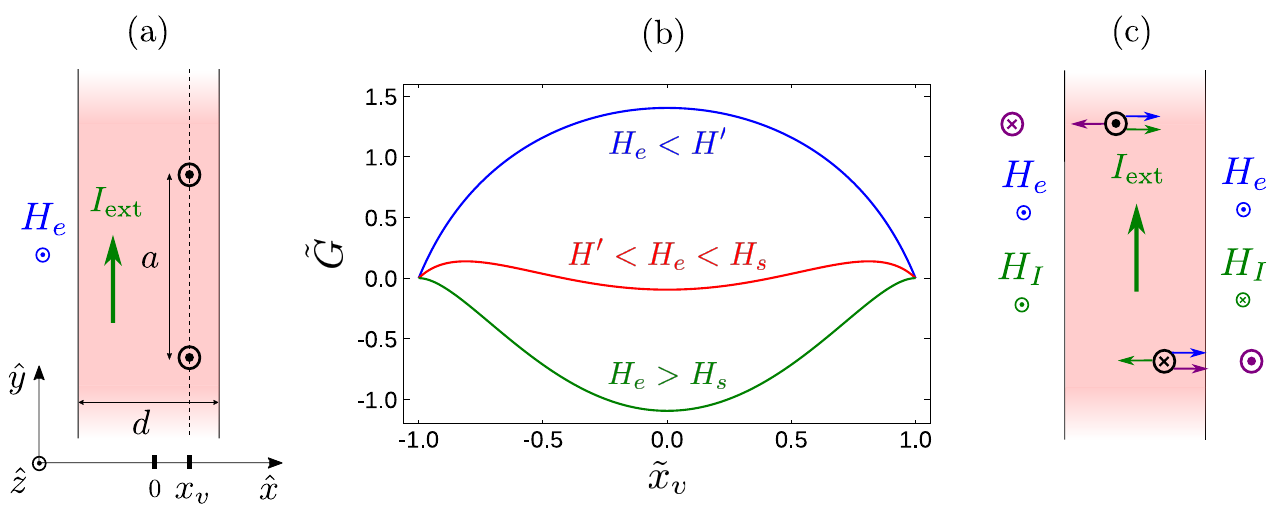}
	\caption{(a) Schematic depiction of the system under consideration. An external magnetic field $H_e ~\hat{\pmb{z}}$ is applied and a current $I_{\mathrm{ext}} ~\hat{\pmb{y}}$ is injected. We consider the superconductor film to host a chain of vortices with flux $\phi_0 ~\hat{\pmb{z}}$, with $\phi_0$ the flux quantum. (b) Normalized Gibbs free energy density $\tilde{G} \equiv G_{1} 4 \pi a d \mu_0 \lambda^2/(V_S \phi_0^2)$ [Eq.~\eqref{eq:g1byvs}] vs. the vortex position $\tilde{x}_v \equiv 2 x_v/d$. In this work, we focus on low fields $H_e < H^\prime$ for which the energy profile does not harbor stable or metastable vortex states in the superconductor. (c) Schematic depiction of the forces experienced by the upper vortex with flux along $\hat{\pmb{z}}$ situated close to the left surface and the lower vortex with flux along $- \hat{\pmb{z}}$ located near the right surface. The force on each vortex is comprised of attraction towards an image antivortex (purple), interaction with the Meissner currents resulting from the applied field (blue), and the Lorentz force due to externally injected current (green). The applied field along $\hat{\pmb{z}}$ lowers the surface barrier experienced by the upper vortex close to the left, while it increases the surface barrier experienced by the lower vortex on the right.}
	\label{fig:SwoSOC}
\end{figure}

We begin by analyzing the vortex instability and critical current in a superconductor film without SOC in this section. We reproduce the key aspects of Shmidt's analysis~\cite{Shmidt1970} and extend it to include vortices with arbitrarily oriented axes, thereby formulating the generalized criterion introduced above [Eq.~\eqref{eq:criterion}]. This generalization is crucial for adequately describing the critical current. 

We consider an infinite superconductor film with thickness $d$ and the y-z plane situated in its middle, as depicted in Fig.~\ref{fig:SwoSOC} (a). An external magnetic field $H_e \hat{\pmb{z}}$ is applied and a current $I_{\mathrm{ext}}$ is injected into the film along $\hat{\pmb{y}}$ via an external circuit. Vortices form regular patterns in their stable states in type II superconductors~\cite{Abrikosov1957,Abrikosov1964}. Thus, we assume the superconductor film to host vortices with flux along $\hat{\pmb{z}}$ located at $x = x_v$ and $y = m a$, where $m$ takes integer values and $a$ is the inter-vortex spacing [see Fig.~\ref{fig:SwoSOC} (a)]. 

The magnetic flux density in the superconductor is evaluated by solving Eq.~\eqref{eq:bmain} with the appropriate boundary conditions and $\pmb{\alpha} = \pmb{0}$. As Eq.~\eqref{eq:bmain} is a linear differential equation, its solution is simply the sum of individual contributions. Thus, the magnetic flux densities contributed by the applied field, the injected current,  and the vortices one at a time add up to yield the total magnetic flux density in the superconductor. The former two contributions are conveniently described by imposing the boundary conditions: $\pmb{B}(x = \pm d/2) = \mu_0 (H_e \mp H_I)~\hat{\pmb{z}}$ [see Fig.~\ref{fig:SwoSOC} (c)], where $H_I$ is the magnetic field generated by the injected current at the superconductor surfaces. The magnetic flux density due to the vortices stems directly from the source terms on the right hand side of Eq.~\eqref{eq:bmain}. Relegating details to appendix \ref{sec:app1}, the resulting magnetic flux density is evaluated as:
\begin{align}\label{eq:bcalc1}
\pmb{B}(x,y) & = \left[ \mu_0 H_e\frac{\mathrm{cosh}\left(\frac{x}{\lambda}\right)}{\mathrm{cosh}\left(\frac{d}{2\lambda}\right)} - \mu_0 H_I\frac{\mathrm{sinh}\left(\frac{x}{\lambda}\right)}{\mathrm{sinh}\left(\frac{d}{2\lambda}\right)} + B_v(x,y) \right] \hat{\pmb{z}},
\end{align}
where $B_{v}(x,y)$ is the vortex chain contribution detailed in appendix \ref{sec:app1}. $H_I$ is determined by requiring that the injected current density $\pmb{j}_{\mathrm{ext}}(x)$ obtained by employing Maxwell's equation on the corresponding contribution to the magnetic flux density [Eq.~\eqref{eq:bcalc1}]:
\begin{align}\label{eq:jext1}
\pmb{j}_{\mathrm{ext}}(x) & = - \frac{\partial}{\partial x} \left( -  H_I\frac{\mathrm{sinh}\left(\frac{x}{\lambda}\right)}{\mathrm{sinh}\left(\frac{d}{2\lambda}\right)} \right) ~\hat{\pmb{y}} = \frac{H_I}{\lambda} \frac{\mathrm{cosh}\left(\frac{x}{\lambda}\right)}{\mathrm{sinh}\left(\frac{d}{2\lambda}\right)} ~ \hat{\pmb{y}},
\end{align}
integrated over area becomes the total injected current:
\begin{align}
\int_{-\frac{d}{2}}^{\frac{d}{2}} dx ~  \pmb{j}_{\mathrm{ext}}(x) \cdot \hat{\pmb{y}} ~ W   & =  I_{\mathrm{ext}}, \\
\implies H_{I} & = \frac{I_{\mathrm{ext}}}{2 W}, \label{eq:I_HI} 
\end{align}
where $W$ is the film width along the z-direction. It is convenient to analyze our system in terms of $H_I$ keeping in mind its one-to-one relation with the externally injected current [Eq.~\eqref{eq:I_HI}].

Employing Eq.~\eqref{eq:bcalc1} in Eq.~\eqref{eq:gibbs1}, we obtain the Gibbs free energy density for the configuration under consideration:
\begin{align}\label{eq:g1byvs}
\frac{G_{1}}{V_S} & = \frac{G_0}{V_S} +\frac{\phi_0 H_e}{ad}\frac{\mathrm{cosh}\left(\frac{x_v}{\lambda}\right)}{\mathrm{cosh}\left(\frac{d}{2\lambda}\right)}  - \frac{1}{ad}\frac{\phi_0^2}{4\pi\mu_0\lambda^2}\left[I_1(x_v)- I_2(x_v)\right],
\end{align}
where $G_0/V_S$ is a constant detailed in appendix \ref{sec:app1} and does not depend on the vortex positions. The term involving $H_e$ captures the interaction between vortices and the Meissner currents generated in order to screen the external field. The term involving $I_{1}(x_v)$, detailed in appendix \ref{sec:app1}, captures the interaction of vortices with the surfaces. It is responsible for Bean-Livingston surface barrier posed to the vortices. The term involving $I_2(x_v)$, also detailed in appendix \ref{sec:app1}, accounts for interaction between the vortices which weakly depends on their distance from the surfaces. As detailed in appendix \ref{sec:app1}, this contribution becomes negligible when $a \gtrsim d$, which is the case of our interest as discussed below. Hence, inter-vortex interaction is disregarded in the rest of our analysis and conveniently, the stability criterion determining the critical current can be formulated considering forces on an individual vortex.

The Gibbs free energy density [Eq.~\eqref{eq:g1byvs}] is plotted against vortex position $x_v$ in Fig.~\ref{fig:SwoSOC}~(b) for different values of the external applied field. At low fields, the energy profile is dominated by the $I_{1}(x_v)$ term, and thus the Bean-Livingston barrier, which prevents the vortices from entering the superconductor. This can also be visualized as an attraction between the vortex and an imaginary image vortex with reverse polarity, as depicted in Fig.~\ref{fig:SwoSOC} (c). For $H_e > H^\prime$, the energy profile develops a minimum at $x_v = 0$ [Fig.~\ref{fig:SwoSOC} (b)]. Vortices are still prevented from entering by a finite surface barrier. But if they do manage to enter somehow, they could find a metastable position in this minimum at $x_v = 0$. $H^\prime$ is determined by the condition:
\begin{align}\label{eq:hprime}
\left. \frac{\partial^2 G_1}{\partial x_v^2} \right|_{x_v = 0} & = 0 ~ \implies H^\prime \approx \frac{2 \phi_0}{\pi \mu_0 d^2}. 
\end{align}
When the external field reaches a larger value $H_s$, the surface barrier is eliminated [Fig.~\ref{fig:SwoSOC}~(b)]. Vortices can freely enter the sample and organize into a regular stable pattern in the middle. $H_s$ is thus defined via:
\begin{align}\label{eq:hs}
\left. \frac{\partial G_1}{\partial x_v} \right|_{x_v = \pm d/2} & = 0 ~ \implies H_s \approx \frac{\phi_0}{2 \pi \mu_0 d \xi}.
\end{align}
It is the high-field regime that Shmidt was primarily interested in. He predicted a ``peak effect'' in the critical current vs. applied field curve that results from an increased stabilization of the vortices at higher fields~\cite{Shmidt1970}. The prediction has been verified experimentally (see Refs.~\cite{Stejic1994,Mawatari1994} and the references therein). Because of this interest in the high field regime, Shmidt only considered vortices whose flux was parallel to the applied field.

In contrast, we limit ourselves to the low-field regime $H_e < H^\prime$ such that no stable or metastable states of vortices exist in the superconductor. As a result, our formulated criterion for critical current [Eq.~\eqref{eq:criterion}] needs to consider forces at the surfaces only, and not within the film. However, we now need to consider all possible directions for the vortex axis since it is not fixed by the external field. Thus, in our considerations, the critical current is the lowest value at which the net force on a vortex at any of the surfaces vanishes for any direction of the vortex axis [Eq.~\eqref{eq:criterion}]. Furthermore, as we consider low fields, the inter-vortex spacing $a$ between an assumed vortex chain is large and we may safely ignore the inter-vortex interactions.

Including the Lorentz force due to injected current in our ongoing analysis of vortices with flux along $\hat{\pmb{z}}$, the critical current is determined using Eq.~\eqref{eq:fnet}:
\begin{align}
- \frac{1}{V_S}  \left. \frac{\partial G_1}{\partial x_v} \right|_{x_v = -d/2} +  \frac{\phi_0 j_{\mathrm{ext}}(-d/2)}{a d} & = 0,
\end{align} 
where $j_{\mathrm{ext}}(x)$ is given by Eq.~\eqref{eq:jext1} and $G_1$ by Eq.~\eqref{eq:g1byvs}. As detailed in appendix \ref{sec:app1}, the equation above yields the critical current expressed via the magnetic field it produces at the surface [Eq.~\eqref{eq:I_HI}]:
\begin{align}\label{eq:hic1}
H_{I_c} & = \frac{d^2}{4 \lambda^2} \left(H_s - H_e\right),
\end{align} 
where $H_s$ is the field at which the Bean-Livingston barrier would be annihilated and is given by Eq.~\eqref{eq:hs} for the case under consideration.

Consistent with Shmidt's analysis for relatively large applied magnetic field along $\hat{\pmb{z}}$, we have so far considered vortex axis to be along $\hat{\pmb{z}}$. In this case, the Lorentz force drives the vortices along $\hat{\pmb{x}}$. Thus, just above the critical current, vortices nucleated on the left surface traverse through the film and annihilate on the right edge [Fig.~\ref{fig:SwoSOC}~(c)]. Consequently, the vortex surface barrier on the left determines critical current. As noted above, since we consider the external field to be weak, we need to consider instability for any vortex axis direction. 

To appreciate this point, let us first consider $H_e = 0$. As depicted in Fig.~\ref{fig:SwoSOC}~(c), the field generated by the injected current points along $\hat{\pmb{z}}$ ($- \hat{\pmb{z}}$) at $x = -d/2$ ($x = d/2$). Thus, two independent vortex instabilities start to develop at each of the surfaces as injected transport current is increased. Vortices with flux along $\hat{\pmb{z}}$ ($- \hat{\pmb{z}}$) are being pushed from left (right) to right (left) and the corresponding instability will be determined by the vortex barrier on the left (right) surface. For equivalent surfaces, the two instabilities will set off at the same critical current. However, two surfaces are never really identical. In practice, one of the two surfaces will have a lower vortex barrier and will determine the critical current. A nonzero $H_e$ applied along $\hat{\pmb{z}}$, helps to reach the instability on the left surface sooner, as clarified via the forces depicted in Fig.~\ref{fig:SwoSOC}~(c). Altogether, the critical current ($\geq 0$) is determined via:
\begin{align}\label{eq:hic2}
H_{I_c} & = \frac{d^2}{4 \lambda^2} \mathrm{min} \left\{ \left(H_{sL} - H_e\right) , \left(H_{sR} + H_e\right)  \right\},  
\end{align}  
where $H_{sL}$ ($H_{sR}$) is the magnetic field at which the vortex surface barrier on the left (right) surface is extinguished. While within our simple model it was evaluated as given in Eq.~\eqref{eq:hs}, the critical current expression above [Eq.~\eqref{eq:hic2}] is general even when $H_{sL,sR}$ are no longer given via Eq.~\eqref{eq:hs}. Due to various reasons, such as disorder and localized enhanced field strengths around corners, the effective value of $H_{s}$ is expected to be lower than what is evaluated in Eq.~\eqref{eq:hs} for an ideal surface~\cite{Shmidt1970,deGennes1966}.  

Thus, generalizing Shmidt's analysis, we have understood the key factors underlying the vortex instability and critical current in the type II superconductor film. From the expression obtained [Eq.~\eqref{eq:hic2}], we can see that the critical current is determined by the weaker of the two surface barriers. We can also anticipate that altering the barriers $H_{sL,sR}$ via a gate voltage, for example, would enable a control over the critical current.     

\section{Superconductor film with SOC and exchange field}\label{sec:SwSOCZ}

Now we include a finite $\pmb{\alpha}$ [Eq.~\ref{eq:GL_free_energy}] in our considerations. As noted before, this contribution to the free energy results from the combined effect of Rashba SOC and an {exchange} spin-splitting field. It is important that the latter is not caused by an external magnetic field, that would necessarily cause an additional orbital contribution, but by an {exchange} interaction~\cite{Mironov2017,Bergeret2018,Miao2015,Hao1991,Moodera2007,Tokuyasu1988,Buzdin2005}. Thus, we consider $\pmb{\alpha}$ in our superconductor to originate in the interfaces with magnetic insulators (MIs), which could be ferromagnets~\cite{Bergeret2018,Hao1991,Eschrig2015,Tokuyasu1988} or antiferromagnets with uncompensated surfaces~\cite{Kamra2018}. Furthermore, we assume $\pmb{\alpha}$ to be small and evaluate its influence on the system up to the first order.

\begin{figure}[tb]
	\centering
	\includegraphics[width=140mm]{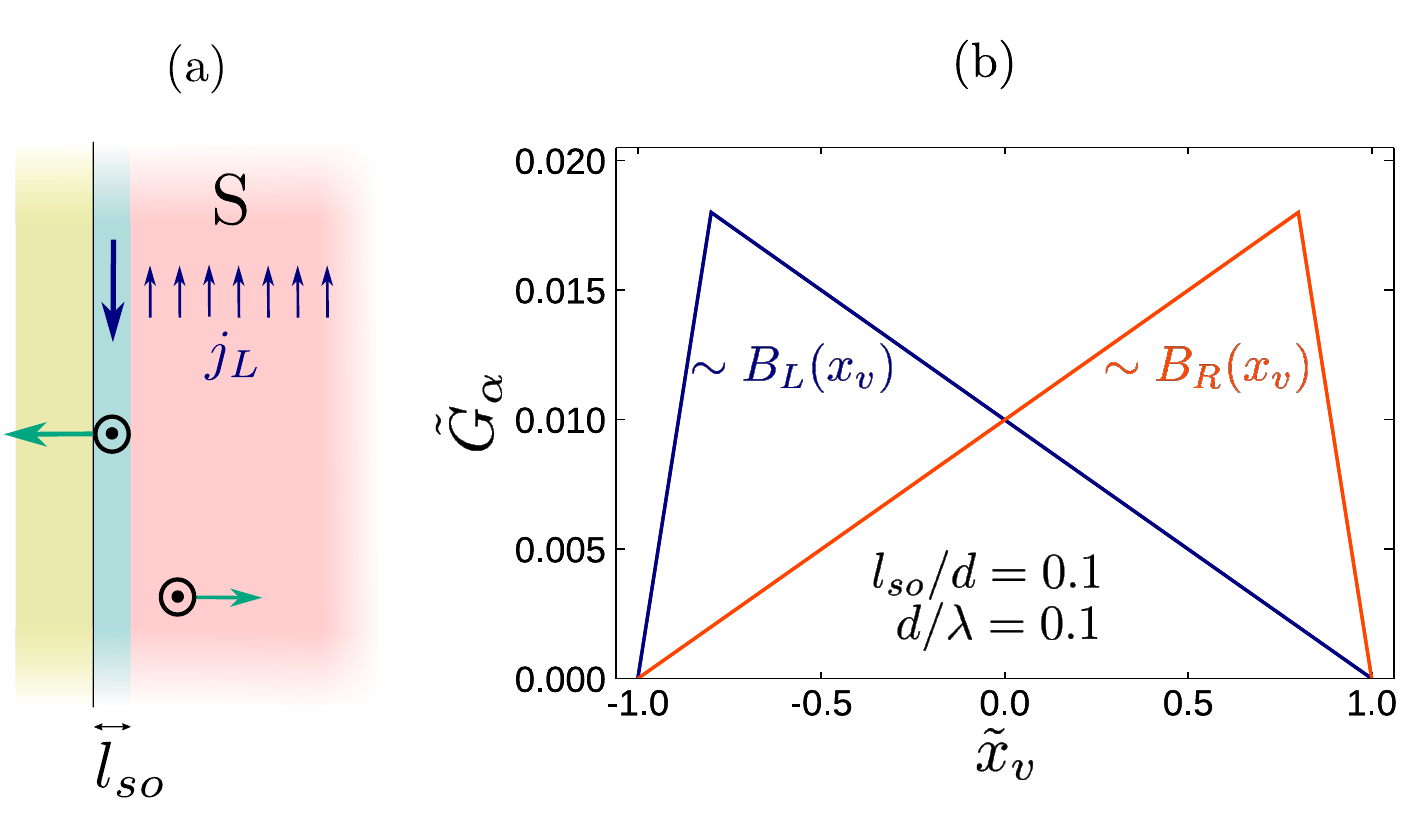}
	\caption{(a) Schematic depiction of spontaneous supercurrent density $\pmb{j}_L$ {(blue arrows)} and {the resulting} forces {(green arrows)} on a vortex {located at different places} in the superconductor. The combined effect of SOC and {exchange} field results in a finite $\pmb{\alpha}$ and {consequently, a} spontaneous supercurrent {directed parallel to $\pmb{\alpha}$ [Eq.~\eqref{eq:jgen}] in the interfacial layer, shaded light blue}. A much smaller oppositely directed supercurrent density throughout the film {(depicted via small blue arrows pointing upwards)} ensures zero net current in equilibrium. (b) The normalized  finite-$\pmb{\alpha}$ Gibbs energy density contributions $\tilde{G}_{\alpha} = G_{\alpha} 4 e \lambda d a \mu_0 / (V_S \phi_0 \alpha_{L,R} \cos \theta_{L,R})$ vs. the normalized vortex position $\tilde{x}_v = 2 x_v / d$. The blue and red curves respectively depict the contributions from left and right interfaces.}
	\label{fig:SwSOC}
\end{figure}

As detailed in Sec.~\ref{sec:overview}, we model the two interfaces with MIs via effective interfacial layers of thicknesses $l_{so}$ that harbor finite $\pmb{\alpha}$ [see Fig.~\ref{fig:SwSOC} (a)]:
\begin{align}
\pmb{\alpha}(x)  = &~ \alpha_L(\mathrm{cos}\theta_L \hat{\pmb{y}} + \mathrm{sin} \theta_L \hat{\pmb{z}}) ~ \Theta(-x-x_{so}) \nonumber \\
  & - \alpha_R(\mathrm{cos}\theta_R \hat{\pmb{y}} + \mathrm{sin} \theta_R \hat{\pmb{z}} ) ~ \Theta(x-x_{so});\quad x \in [-d/2, d/2], \label{eq:alpha_definition}
\end{align}
where $\alpha_{L,R}$ parameterize $\pmb{\alpha}$ magnitude at the left and right interfaces, $\theta_{L,R}$ correspondingly allow for their general in-plane directions, $x_{so} \equiv d/2 - l_{so}$, and $\Theta(x)$ is the heaviside step function. We now need to solve Eq.~\eqref{eq:bmain} to obtain the magnetic flux density in the superconductor. Since this is a linear equation, the contribution of finite $\pmb{\alpha}$ simply adds to the magnetic flux density presented in the previous section for $\pmb{\alpha} = \pmb{0}$ [Eq.~\eqref{eq:bcalc1}], and is evaluated as:
\begin{align}\label{eq:balpha}
\pmb{B}_{\alpha}(x) = & -[\mathrm{cos}\theta_L B_L(x) + \mathrm{cos}\theta_R B_R(x)] ~ \pmb{\hat{z}}  + [\mathrm{sin}\theta_L B_L(x) + \mathrm{sin}\theta_R B_R(x)]~ \pmb{\hat{y}},
\end{align} 
where 
\begin{align}
B_{L}(x) & = \frac{\alpha_{L}}{4e\lambda} \frac{\mathrm{cosh}\left(\frac{d-|x+x_{so}|}{\lambda}\right) - \mathrm{cosh}\left( \frac{x-x_{so}}{\lambda} \right)}{\mathrm{sinh}\left( \frac{d}{\lambda} \right)} \label{eq:B_L}, \\
B_R(x) & = \frac{\alpha_R}{4e\lambda} \frac{\mathrm{cosh}\left(\frac{d-|x-x_{so}|}{\lambda}\right) - \mathrm{cosh}\left( \frac{x+x_{so}}{\lambda} \right)}{\mathrm{sinh}\left( \frac{d}{\lambda} \right)} \label{eq:B_R}.
\end{align}
Employing the total magnetic flux density in Eq.~\eqref{eq:gibbs1}, we obtain the total average Gibbs free energy density:
\begin{align}\label{eq:gibbswSOC}
\frac{G_2}{V_S} & =  \frac{G_1}{V_S} + \frac{G_{\alpha 0}}{V_S} + \frac{G_{\alpha}}{V_S},
\end{align}
where $G_1$ is the contribution without SOC [Eq.~\eqref{eq:g1byvs}], $G_{\alpha 0}$ does not depend on the vortex position and has been detailed in appendix \ref{sec:app2}, and 
\begin{align}\label{eq:galpha}
\frac{G_{\alpha}}{V_S} & = - \frac{\phi_0}{da\mu_0}\left[ B_L(x_v)\mathrm{cos}\theta_L + B_R(x_v)\mathrm{cos}\theta_R \right],
\end{align}
constitutes the Gibbs energy contribution resulting from finite $\pmb{\alpha}$, that exerts additional forces on the vortices, evaluated via Eq.~\eqref{eq:fnet}.

We pause to physically interpret our mathematical results. For simplicity, let us focus on the effect of the left interface by assuming $\alpha_R = 0$. The contribution of the right interface can be understood in an analogous manner. On account of finite $\alpha_L$, a spontaneous supercurrent density is generated in the interfacial layer with thickness $l_{so}$ next to the interface. Since the total equilibrium current through the film must vanish, this causes a weaker oppositely directed supercurrent density in the entire superconductor layer [see Fig.~\ref{fig:SwSOC} (a)]. Consequently, this spatially dependent spontaneous supercurrent density exerts a relatively large force on a vortex located at the interface and a weak force if it is located in the bulk. This can also be seen via the slope of the Gibbs free energy density contribution due to $\pmb{\alpha}$ plotted in Fig.~\ref{fig:SwSOC} (b). Thus, the spontaneous supercurrents at the interfaces significantly alter the vortex surface barrier fields and consequently the critical current. 

To determine the critical current, we should employ the instability condition Eq.~\eqref{eq:criterion} for all possible vortex axes. However, assuming $\pmb{\alpha}$ to be small, we expect the vortex axis for the instability at each of the interface to be determined primarily by the magnetic flux density generated by the externally injected current. This is analogous to our considerations without SOC in the previous section, where we found the critical current to be determined by the vortex surface barrier at each interface, as given by Eq.~\eqref{eq:hic2}. Following the analogous analysis, the critical current including the effect of SOC can still be expressed via Eq.~\eqref{eq:hic2} with $\pmb{\alpha}$ contributing corrections to the surface barrier fields evaluated to be:
\begin{align}
\Delta H_{sL} & = -\frac{1}{\mu_0ed}\left[\alpha_L\mathrm{cos}\theta_L\left(1-\frac{l_{so}}{d}\right) + \alpha_R\mathrm{cos}\theta_R\left(\frac{l_{so}}{d}-\frac{l_{so}d}{4\lambda^2}\right)\right] \approx -\frac{\alpha_L\mathrm{cos}\theta_L}{\mu_0ed}, \label{eq:deltahsl}  \\
\Delta H_{sR} & = \frac{1}{\mu_0ed}\left[\alpha_R\mathrm{cos}\theta_R\left(1-\frac{l_{so}}{d}\right) + \alpha_L\mathrm{cos}\theta_L\left(\frac{l_{so}}{d}-\frac{l_{so}d}{4\lambda^2}\right)\right] \approx \frac{\alpha_R\mathrm{cos}\theta_R}{\mu_0ed} . \label{eq:deltahsr}
\end{align}
Hence, the vortex surface barrier, or equivalently the critical current, may be increased or decreased due to $\alpha_{L,R}$. They may further be controlled via angles $\theta_{L,R}$, which are determined by the direction of magnetic or N\'eel order in the adjacent MIs.

\section{Superconductor-magnet hybrids}\label{sec:S-MI}

We now examine the critical current in experimentally relevant heterostructures formed by a type II superconductor (S) film adjoining magnetic insulator (MI) film(s) on one or both sides. The latter's role is to provide an effective {exchange} field $\pmb{h}$ that determines $\pmb{\alpha} \propto \alpha_{Ra} \hat{\pmb{n}} \times \pmb{h}$ as discussed in Sec.~\ref{sec:overview}. The field $\pmb{h}$ results from interfacial {exchange} interaction and is collinear with the magnetization of the adjacent ferromagnet or the sublattice-magnetization of the uncompensated antiferromagnet~\cite{Kamra2018}. {Furthermore, recent considerations show that a range of metallic antiferromagnets harbor a strong spin-splitting of their electronic states~\cite{Pekar1964,Ahn2019,Hayami2019,Smejkal2020,Yuan2020,Egorov2021}. These can further provide an effective {exchange} field collinear with the antiferromagnetic N\'eel vector even with a compensated interface~\cite{Johnsen2021}.} Hence, the direction of $\pmb{h}$, and thus $\pmb{\alpha}$, can be conveniently controlled by a weak external magnetic field. In case of a ferromagnet, $\pmb{h}$ simply becomes collinear with the applied field. For a typical easy-plane antiferromagnet, the N\'eel order and $\pmb{h}$ orient perpendicular to the applied field (e.g., see Ref.~\cite{Wimmer2020}). For both cases, the magnitude of $\pmb{h}$ is independent of the external field and is determined by the MI and interface properties. Hence, the critical current in heterostructures of interest can be controlled via the orientation of a weak applied magnetic field. Furthermore, a gate-voltage may control the Rashba SOC parameter $\alpha_{Ra}$~\cite{Nitta1997,Manchon2015,Chen2018}.

In our considerations, we continue to assume the effects of applied magnetic field and SOC to be weak. We evaluate the critical current up to the first order in both these variables. This also implies that the vortex axes for the instability that determines the critical current is governed primarily by the magnetic field generated by the externally injected current at the surfaces of S. The situation is thus the same as our analyses in the previous sections and we may express critical current using Eqs.~\eqref{eq:I_HI} and \eqref{eq:hic2}:
\begin{align}\label{eq:icfinal}
I_{c} = & \frac{2 W d^2}{4 \lambda^2} \mathrm{min} \left\{ \left(H_{sL} - H_e\right) , \left(H_{sR} + H_e\right)  \right\},  
\end{align} 
where $W$ is the film width along z direction (perpendicular to the injected current flow). We further split the surface barrier fields $H_{sL,sR} \equiv H_{sL,sR}^0 + \Delta H_{sL,sR}$ into their values without interfacial SOC ($H_{sL,sR}^0$) plus the difference that can be controlled via external parameters. Similarly, we express the total critical current as $I_c \equiv I_{c}^0 + \Delta I_c$. Equation \eqref{eq:icfinal}, together with the results evaluated in the previous section [Eqs.~\eqref{eq:deltahsl} and \eqref{eq:deltahsr}], forms the basis for our critical current analysis below.

\subsection{MI-S bilayer}

\begin{figure}[tb]
	\centering
	\includegraphics[width=\columnwidth]{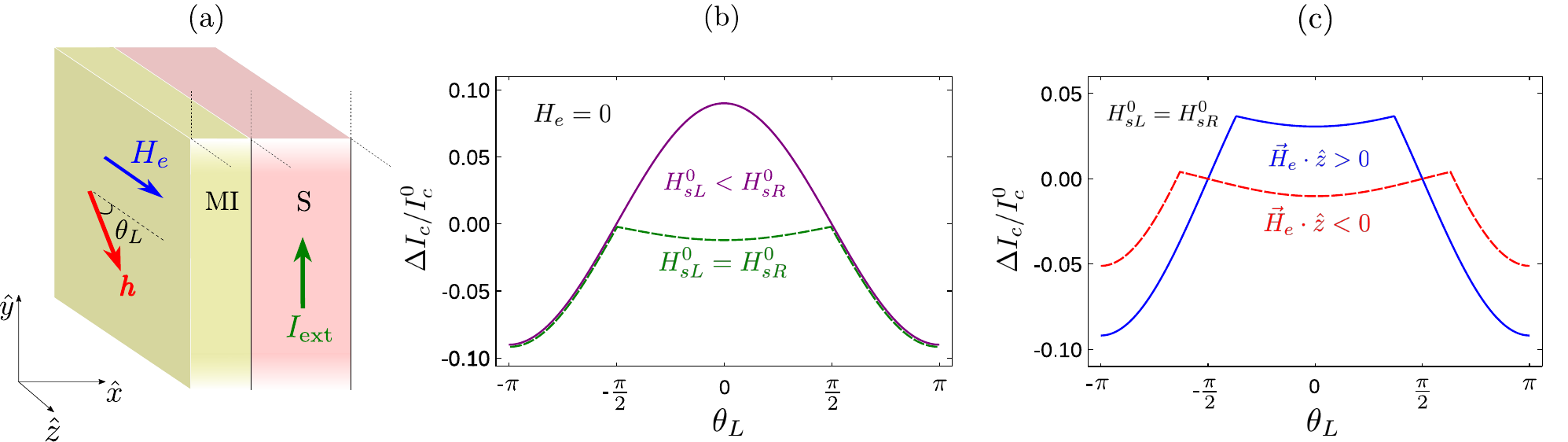}
	\caption{(a) Schematic of the magnetic insulator (MI)- superconductor (S) bilayer that enables a control of the critical supercurrent in S layer. The external current is injected along $\hat{y}$ and the MI magnetic order determines the {exchange} field $\pmb{h}$, which can be controlled by, for example, an external magnetic field. (b) and (c) Change in the superconducting critical current $\Delta I_c$ due to the combined action of {exchange} field and interfacial SOC resulting in a finite $\pmb{\alpha}$ vs. orientation of the {exchange} field $\pmb{h}$. $I_{c}^0$ is the critical current without interfacial SOC. We assume $\alpha_L/ e \mu_0 d H_{sL}^0 = 0.1$. In (c), we consider $H_e/H_{sL}^0 = 0.02$.}
	\label{fig:MIS}
\end{figure}

We first consider a MI-S heterostructure as depicted in Fig.~\ref{fig:MIS} (a), which corresponds to $\alpha_R = 0$. The magnetic order parameter, and thus $\pmb{h}$, is considered to make an angle $\theta_L$ with the z axis, which is defined as perpendicular to the injected current flow. Such a bilayer structure is likely to have unequal vortex surface barrier fields at the two S surfaces due to their different qualities. For example, disorder at the S$|$MI interface is likely to lower $H_{sL}^0$ compared to the right interface terminated in vacuum (or a capping layer). Under such a condition $H_{sL}^0 < H_{sR}^0$, the critical current is primarily determined by the left interface [Eq.~\eqref{eq:deltahsl}] and is depicted via solid purple line in Fig.~\ref{fig:MIS} (b). Thus, SOC-mediated change in the critical current can be positive or negative, depending on $\theta_L$ and the sign of $\alpha_L$. When we consider identical surface barriers such that $H_{sL}^0 = H_{sR}^0$, the critical current varies as depicted by the green dashed curve in Fig.~\ref{fig:MIS} (b). In this case, for a range of $\theta_L$, the vortex instability is determined by the right surface [Eq.~\eqref{eq:deltahsr}]. It depends weakly on $\alpha_L$ due to the supercurrent density ($\sim l_{so}/d$) generated throughout S to cancel the spontaneous supercurrent induced at the left interface. The range of $\theta_L$ for which the critical current is determined by the left interface shows a decrease compared to the value without SOC. Thus, SOC contribution in this case may only lower the critical current compared to $I_{c}^0$. In Fig.~\ref{fig:MIS} (c), we depict the case of $H_{sL}^0 = H_{sR}^0$ accounting for the small correction that results from a weak but finite external magnetic field along the z axis. This quantifies how the applied field lowers the vortex barrier at one surface while raising it on the other, as detailed in Sec.~\ref{sec:SwoSOC} and adequately captured in Eq.~\eqref{eq:icfinal}.  

When we consider $H_{sL}^0 > H_{sR}^0$, critical current is determined by the vortex instability at the right surface and is practically unaffected by external parameters that can influence $\alpha_L$. Thus, in order to accomplish MI-S heterostructures admitting a control of the critical supercurrent, we must pay special attention to having a high quality of the other S surface, such that $H_{sL}^0 < H_{sR}^0$. This might occur naturally if S is deposited on a MI layer and left exposed to vacuum or a capping layer.

\subsection{MI-S-MI trilayer}

\begin{figure}[tb]
	\centering
	\includegraphics[width=\columnwidth]{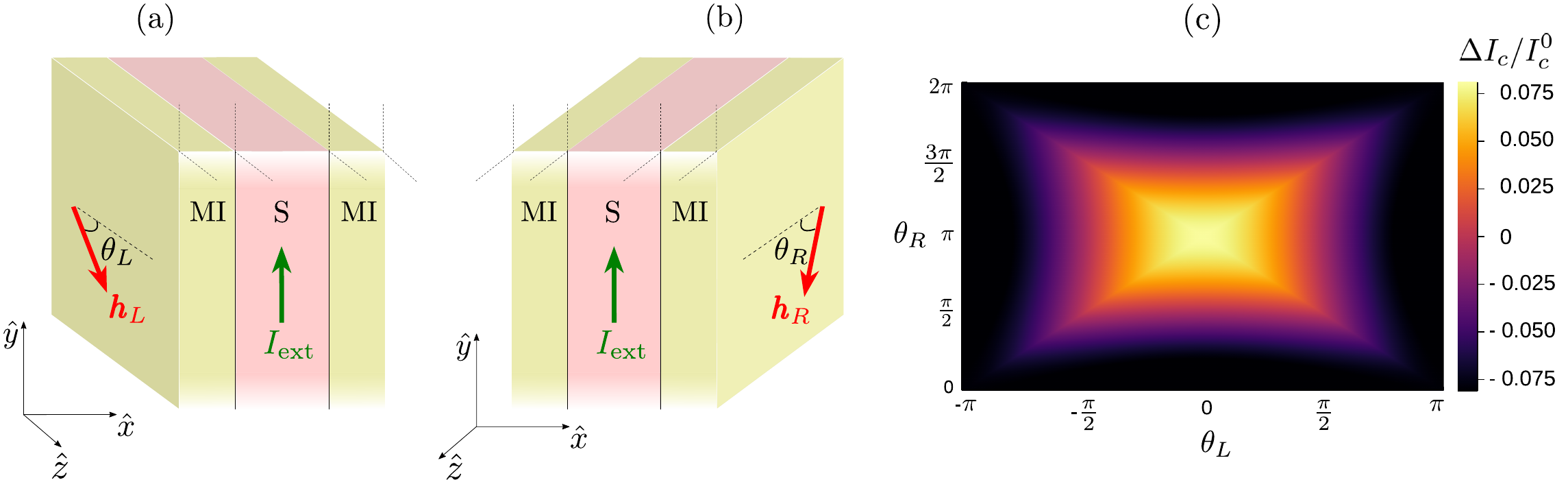}
	\caption{(a) and (b) Schematic depiction of the trilayer device that allows a control of critical current in S layer via magnetic orders in the two adjacent MIs. (c) Fractional change in the critical current vs. orientations of the {exchange} fields induced by the two adjacent MIs. We assume $H_e = 0$, $H_{sL}^0 = H_{sR}^0$, $|\alpha_L| = |\alpha_R|$, and $\alpha_{L}/ e \mu_0 d H_{sL}^0 = 0.1$.}
	\label{fig:MISMI}
\end{figure}

Considering the S layer to be sandwiched between two MI layers, as depicted in Fig.~\ref{fig:MISMI}, provides additional control. This trilayer structure further allows to witness a full interplay between the vortex barriers at both the surfaces and makes the parameter space larger with numerous possibilities. The critical current is still governed by the relatively simple result Eq.~\eqref{eq:icfinal} and can be obtained for any desired parameter space. Considering $H_e = 0$ and symmetrical vortex surface barriers in the absence of SOC i.e., $H_{sL}^0 = H_{sR}^0$, the critical current varies with the MI orders as depicted in Fig.~\ref{fig:MISMI} (c). Thus, we find the highest critical current when the two order parameters are antiparallel assuming the same sign of the Rashba SOC at the two interfaces, which can further be controlled via gate voltage(s) using additional electrode layers not shown explicitly in Fig.~\ref{fig:MISMI}. In practice, one might consider the magnetic order in one of the MIs to be fixed, for example by {exchange}-biasing~\cite{Nogues1999}, and the other MI then allows $I_c$ to be controlled via a weak applied magnetic field. The generality of our result Eq.~\eqref{eq:icfinal} should facilitate engineering different means of controlling the superconducting critical current in such superconductor-magnet hybrids~\cite{Li2013,Moraru2006,Zhu2017}.

\section{Superconductor film with gate voltage-controlled SOC}\label{sec:SwSOC}

Since the present work is partly inspired by the recent observation of gate voltage-induced critical current enhancement in superconducting NbN~\cite{Rocci2021}, we pause to discuss the implications of our theory on these experimental results. The experiments in question recorded the critical current $I_c$ of a NbN bridge deposited on insulating Si$\mathrm{O}_2$ and gated via the p-doped Si substrate. NbN layer was thus sandwiched between two non-magnetic insulators and no magnetic field was applied. The critical current was found to increase with the applied gate voltage (almost) irrespective of the latter's polarity. A small asymmetry with respect to the gate voltage sign was, however, observed.

Our analysis above requires an MI-induced {exchange} field to achieve changes in the critical current. This is because we have retained terms up to the first order in the interfacial Rashba SOC $\alpha_{\mathrm{Ra}} \propto \pmb{\alpha} $. In the absence of {exchange} field from an adjacent MI, the leading order effect of the interfacial SOC on the vortex surface barrier is expected to scale as $\alpha_{\mathrm{Ra}}^2$~\cite{Silaev2020}. Evaluating this within an analytic model seems difficult, as discussed further in Sec.~\ref{sec:discussion} below. Nevertheless, we proceed to understand the symmetries of critical current modulation induced by such an effect. Without loss of generality, we may assume that $H_{sL} < H_{sR}$ such that the critical current is given via [Eq.~\eqref{eq:icfinal}]:
\begin{align}
I_{c} & = \frac{2 W d^2}{4 \lambda^2} H_{sL}, \label {eq:ichsl}\\
\frac{\Delta I_c}{I_{c}^0} & = \frac{\Delta H_{sL}}{H_{sL}^0}  \propto \alpha_{\mathrm{Ra}}^2 \ = \alpha_0^2 + c_1^2 V_G^2 + 2 c_1 \alpha_0 V_G, \label{eq:expic}
\end{align} 
where $I_{c}^0$ ($\Delta I_c$) and $H_{sL}^0$ ($\Delta H_{sL}$) denote the critical current and the surface barrier field without (due to) the interfacial Rashba SOC. We have further assumed $\alpha_{\mathrm{Ra}} = \alpha_0 + c_1 V_G$, with $V_G$ the gate voltage~\cite{Chen2018}. Equation \eqref{eq:expic} therefore clarifies the predominantly bipolar (even in $V_G$) nature of the critical current dependence on the applied gate voltage with a possible polarity-dependence resulting from nonzero $\alpha_0$. Further, the mechanism under investigation does not influence the superconducting Tc, consistent with the experiments~\cite{Rocci2021}.

\section{Estimation of the critical current modulation}\label{sec:estimation}
In this section, we estimate the expected critical current modulation in the S-MI hybrids considered above. A change in the critical current can be accomplished via the MI order (e.g., magnetization of a ferromagnet) orientation or an applied gate voltage that alters the interfacial Rashba SOC strength.

As per our discussion in the previous section, the fractional change in the critical current may be expressed as:
\begin{align}
\left| \frac{\Delta I_c}{I_{c}^0} \right| & = \left| \frac{\Delta H_{sL}}{H_{sL}^0} \right|, \\
  & = \left| \frac{\alpha_L \mathrm{cos}\theta_L}{\mu_0 e d H_{sL}^0} \right|, \label{eq:estimate1}
\end{align}
where we have employed Eq.~\eqref{eq:deltahsl} in achieving the above simplification. We can further substitute for $H_{sL}^0$ using Eq.~\eqref{eq:hs}, but that is unlikely to reflect the situation in samples of interest. The surface barrier field as derived in Eq.~\eqref{eq:hs} for an ideal surface is rather large. If it were to determine the critical current, the latter will be comparable to the pair-breaking mechanism-determined critical current and vortex instability mechanism may not be operational at all. As discussed previously, in realistic samples, interfacial disorder and magnetic field inhomogenieties around the edges substantially reduce $H_{sL}^0$ making the vortex instability the dominant mechanism in a range of samples~\cite{Shmidt1970,deGennes1966,Mawatari1994}. Thus, we estimate $H_{sL}^0$ using Eq.~\eqref{eq:ichsl} and the experimentally measured value of $I_{c}^0$. Expressing $\alpha_L$ in terms of the induced {exchange} field magnitude $h$ and dropping the subscript $L$ denoting the left interface, we obtain [Eq.~\eqref{eq:estimate1}]:
\begin{align}\label{eq:estimate2}
\left| \frac{\Delta I_c}{I_{c}^0} \right| & = \left| \frac{ \beta h \mathrm{cos}\theta}{\mu_0 e d H_{s}^0} \right|.
\end{align}
The factor $\beta h$ can be expressed in terms of the Rashba parameter $\alpha_{\mathrm{Ra}}$ and properties of the superconductor via~\cite{Olthof2019,Dimitrova2007}:
\begin{align}\label{eq:estimate3}
\beta h & = \left( \frac{v_{\mathrm{Ra}}}{v_F} \right) \left( \frac{c}{l_{so}} \right) \left( \frac{h}{k_B T_c} \right) \frac{\hbar}{\xi},
\end{align}
where $v_{\mathrm{Ra}} = \alpha_{\mathrm{Ra}}/\hbar$ is the Rashba velocity, $v_F$ is the Fermi velocity, $c$ is the lattice constant, $T_c$ is the superconducting critical temperature, and $h$ is the {exchange} field in units of energy. In the equation above, the parentheses enclose dimensionless quantities. Employing Eq.~\eqref{eq:estimate3} in Eq.~\eqref{eq:estimate2} and dropping the explicit $\theta$ dependence, we obtain:
\begin{align}\label{eq:estimatefin}
\left| \frac{\Delta I_c}{I_{c}^0} \right| & = \left( \frac{v_{\mathrm{Ra}}}{v_F} \right) \left( \frac{c}{l_{so}} \right) \left( \frac{h}{k_B T_c} \right) \left( \frac{\phi_0}{ \pi d \xi \mu_0 H_{s}^0} \right),
\end{align}
where we continue to enclose dimensionless quantities in parentheses. Equation \eqref{eq:estimatefin} is our desired estimation and sheds light on the key features that we pause to discuss. 

The first parenthesis on the right hand side of Eq.~\eqref{eq:estimatefin} encloses the ratio between SOC energy splitting at the Fermi wavevector and the Fermi energy. This quantity can be tuned in a wide range via an applied gate voltage for different systems~\cite{Manchon2015,Caviglia2010,Ben2010,Nitta1997,Chen2018}. The second parenthesis encloses the inverse of the number of monolayers in the length $l_{so}$, the value of which has been estimated differently in different works~\cite{Mironov2017,Olthof2019,Dimitrova2007}. We expect $l_{so}$ to be roughly the same as the coherence length $\xi$. The third paranthesis encloses the {exchange} field as a fraction of the critical temperature. All the three preceding factors are typically less than 1. The fourth and final factor is the ratio between the magnetic flux density obtained if the flux quantum was distributed over an area $d \xi$ to the magnetic flux density corresponding to the vortex surface barrier field of the superconducting film without SOC. This final factor can be much larger than 1. Thus, to maximize the critical current modulation, we should employ superconductors with small coherence lengths and film thicknesses.

In order to obtain a numerical estimate for the critical current modulation, we consider the NbN device investigated in the recent experiments~\cite{Rocci2021}. Thus, we consider $d = 10$ nm, $\xi = 10$ nm, $\lambda = 450$ nm, $I_{c}^0 = 80~\mu$A and $W = 1~\mu$m to obtain $\mu_0 H_{s}^0 \approx 0.4$ T. Further, assuming the SOC splitting to be $5\%$ of the Fermi energy and a lattice constant of $0.4$ nm~\cite{Shiino2010}, we estimate [Eq.~\ref{eq:estimatefin}]:
\begin{align}
\left| \frac{\Delta I_c}{I_{c}^0} \right| & = \left( 0.05 \right) \left( 0.04 \right) \left( 0.2 \right) \left( 16 \right)  = 0.006,
\end{align}
where we considered $l_{so} = \xi$. This represents a modest modulation of about $1\%$ that can be increased substantially via careful engineering of devices.

\section{Discussion and Outlook}\label{sec:discussion}

{The main contribution of our analysis can be divided into two fairly independent parts. The first is establishing a relation between the critical current and the vortex surface barriers. This is expressed in a rather general form by Eq.~\eqref{eq:criterion}, and for a somewhat specialized case by Eq.~\eqref{eq:hic2}. The general idea is that if we know the vortex surface barrier for all surfaces and flux orientations, we may evaluate the critical current due to this mechanism. If it is found to be smaller than the critical current due to all other mechanisms, the film will be governed by this mechanism and we may control its critical current via the surface barrier. The second main part of our work has been evaluating the vortex surface barriers using simplified analytically tractable models. While our considerations clearly demonstrate an SOC-enabled control over vortex surface barriers, the latter are generally difficult to evaluate for realistic samples~\cite{Benfenati2020} and should be treated as experimental unknowns, where the theory may provide guidance. However, our finding that the interfacial Rashba SOC should affect the vortex surface barrier is clear and independent of the detailed modeling.}

{With this in mind, let us examine when our investigated mechanism of critical current should dominate realistic samples. In a disordered type II superconductor film, the vortices (or flux) get pinned in the bulk, thereby eliminating the role of interfacial barriers. Thus, clean films devoid of flux pinning centers are required. Due to the same reason, thin clean films are better, since there is then a small distance to be traversed between the surfaces. Thicker films will have a higher chance of vortices encountering flux pinning centers while moving from one surface to the other. Further, within our analytic model, the evaluated surface barrier [Eq.~\eqref{eq:hs}] is too high for it to determine the critical current. However, as noted by Shmidt~\cite{Shmidt1970}, realistic samples could manifest much weaker surface barriers due to various reasons, such as interfacial disorder. Hence, clean thin films grown on a substrate are good candidates, and the vortex surface barrier is best treated as an experimental unknown.} 

{Furthermore, vortices with out-of-plane flux may enter the superconducting film from other surfaces not explicitly considered in our study. These might even face a smaller surface barrier due to the large quasi-bulk distance between the opposite surfaces. However, these will suffer from the challenge of flux-pinning in the bulk since they have to travel a long distance between the two surfaces. Also, the magnetic field generated by the current flowing through the film favors spontaneous nucleation of vortices with in-plane flux. A more careful consideration of the out-of-plane vortices requires a numerical three-dimensional analysis and is left for a future study.}

We now note some weaknesses of our analysis and suggest ways to address them in future works. We have worked in the limit $\xi \ll d \ll \lambda$ and found that the highest critical current modulation is achieved for the smallest $d$. Thus, we estimated the modulation by considering $d = \xi$, which is strictly speaking beyond the applicability of our theory. Simply put, we work in the London limit corresponding to $\xi \to 0$ and often have to choose other lengths comparable to $\xi$. Despite such transgressions, our analysis here offers good estimates as explained further in the appendix~\ref{sec:app1} and also by Shmidt, whose theory faced the same challenges~\cite{Shmidt1970}. While we have captured the qualitative physics adequately, a quantitative reliance on our expressions is not encouraged. 

Furthermore, we assumed the effect of interfacial SOC to be small and evaluated corrections to the critical current perturbatively up to the first order in $\alpha_{\mathrm{Ra}}$. In this sense, we have assumed the critical current modulation to be small at the outset. This allowed us to treat the vortices as robust objects that are not affected by the SOC. If we attempt to evaluate the Gibbs free energy up to the second order in $\alpha_{\mathrm{Ra}}$, we should also account for a distortion in the shape of the vortices as well as corrections to their core energy. This analysis, although analytically tractable, seems discouragingly tedious. Furthermore, the case of highest experimental interest, admitting a large critical current modulation, corresponds to a strong interfacial SOC that would invalidate our perturbative approach altogether and necessitate an exact numerical treatment.

Both the above-mentioned weaknesses of our analysis are good news. Within our proposed mechanism, the experiments can achieve a much stronger critical current modulation than adequately described by our analytical theory. The goal of this work has been to clarify the physics of critical current control by modifying the vortex surface barrier using a simple analytically tractable model. Having understood the key ingredients and qualitative dependencies, the experiments can engineer new hybrids capable of robust critical current control via gate voltage and/or magnetic order orientation. Further, important insights can be achieved using recently established methods for measuring interfacial spin-orbit torques~\cite{Muller2021}. On the theory side, the vortex surface barriers and corresponding critical currents should be evaluated numerically~\cite{Benfenati2020} for various superconducting hybrids of interest~\cite{Mercaldo2020,Olthof2021,Johnsen2020}, for example using the quasiclassical Green's function method~\cite{Belzig1999}. The recent new insights regarding orbital contributions to magnetization in superconductors with Rashba SOC~\cite{Chirolli2021} pose another important question and could offer an avenue for enhancement of the critical current modulation.

\section{Summary}\label{sec:summary}
We have theoretically investigated the effect of interfacial Rashba spin-orbit coupling on the vortex surface barrier in type II superconducting films interfaced with one or more magnetic insulators. By formulating a general criterion for vortex instability, we relate the critical current in such films to the vortex surface barrier. Thus, we predict control of critical current in type II superconducting films by influencing the vortex surface barrier. Experimentally verifiable dependencies of this critical current modulation in different superconductor-magnet hybrids have been worked out. We have also delineated dependencies on the various properties of the hybrid thereby providing design equations for engineering devices with maximal critical current control. Our simple analytic theory presented herein lays the foundation for and would benefit from more detailed numerical works in the future.

\section*{Acknowledgments}
{We thank Emmanuel I. Rashba for valuable discussions.} We acknowledge financial support from the Spanish Ministry for Science and Innovation -- AEI Grant CEX2018-000805-M (through the ``Maria de Maeztu'' Programme for Units of Excellence in R\&D), and the Research Council of Norway through its Centers of Excellence funding scheme, project 262633, ``QuSpin''. Work at MIT is supported by ARO (W911NF-20-2-0061), NSF (DMR 1700137 and NSF C-Accel Track C Grant No. 2040620), ONR (Nos. N00014-16-1-2657 and N00014-20-1-2306), and CIQM-NSF DMR-1231319.

\appendix

\section{Superconductor film without SOC}\label{sec:app1}
Here, we go through the analysis of a superconductor film without spin-orbit coupling in some detail, providing the tedious mathematical expressions that have been avoided in the main text. We consider a superconductor film hosting a chain of vortices as described in Sec.~\ref{sec:SwoSOC}. The magnetic flux density $\pmb{B}(x,y)$ is determined by solving Eq.~\eqref{eq:bmain} which, for the case under consideration, is given by:
\begin{align}\label{eq:Appendix_magnetic_field_equation}
\lambda^2 \nabla^2 \pmb{B} - \pmb{B} & = - \phi_0\sum_m \delta(x-x_v)\delta(y-ma)~ \hat{\pmb{z}}; \nonumber \\ 
\pmb{B}(x = \pm d/2,y) & = \mu_0 (H_e \mp H_I) ~\hat{\pmb{z}}.
\end{align}
As discussed in Sec.~\ref{sec:SwoSOC}, the above boundary condition adequately describes the applied field $H_e~\hat{\pmb{z}}$ and the injected current $I_{\mathrm{ext}}~\hat{\pmb{y}}$. The solution to Eq. \eqref{eq:Appendix_magnetic_field_equation} is~\cite{Abrikosov1964,Shmidt1970}
\begin{equation}\label{eq:Appendix_magnetic_field}
\pmb{B}(x, y) = \left[\mu_0 H_e \frac{\mathrm{cosh}\left(\frac{x}{\lambda}\right)}{\mathrm{cosh}\left(\frac{d}{2\lambda}\right)} - \mu_0 H_I \frac{\mathrm{sinh}\left(\frac{x}{\lambda}\right)}{\mathrm{sinh}\left(\frac{d}{2\lambda}\right)} + \sum_m B_m(x, y)\right] \pmb{\hat{z}},
\end{equation}
where 
\begin{align}\label{eq:Appendix_B_m_and_B_k}
B_m(x, y) & = \int_{-\infty}^{\infty}\frac{d k}{2\pi}e^{ik(y-ma)}B_k(x); \nonumber \\
 B_k(x) & = \frac{\phi_0}{2\lambda^2 u}\frac{\mathrm{cosh}[u(d-|x-x_v|)]-\mathrm{cosh}[u(x+x_v)]}{\mathrm{sinh}(ud)},
\end{align}
and $u \equiv \sqrt{k^2 + 1/\lambda^2}$. 

The average Gibbs free energy density can now be evaluated. To enable comparison with literature~\cite{Abrikosov1964,Shmidt1970}, we include the energy of the vortex core in the following analysis. We further assume that the vortex core energy does not depend on its position and consider it to be the same as in an infinite superconductor. Thus, its inclusion in our analysis does not lead to any additional forces on the vortex and does not influence the evaluated critical current. The resulting average Gibbs free energy density is [Eq.~\eqref{eq:gibbs1}]~\cite{Abrikosov1964,Shmidt1970}
\begin{align}
\frac{G}{V_S} = & \frac{\varepsilon}{ad} - \frac{1}{d}\lim_{Y\to\infty} \frac{1}{Y}\int_{-Y/2}^{Y/2} d y \int_{-\infty}^{\infty}dx\frac{1}{2\mu_0}[\pmb{B}_\infty^2+\lambda^2(\nabla\times \pmb{B}_\infty)^2] \nonumber \\
& + \frac{1}{d}\lim_{Y\to\infty} \frac{1}{Y}\int_{-Y/2}^{Y/2}d y \int_{-d/2}^{d/2}dx \left[ \frac{1}{2\mu_0}[\pmb{B}^2+\lambda^2(\nabla\times \pmb{B})^2] - \pmb{B} \cdot H_e \hat{\pmb{z}} \right],
\label{eq:Appendix_free_energy_Average}
\end{align}
where $\varepsilon$ is the total energy per unit length of an isolated vortex in an infinite superconductor, and $\pmb{B}_\infty$ is its associated magnetic field. The first two terms in Eq.~\eqref{eq:Appendix_free_energy_Average} together represent the vortex core energy of the chain, while the remaining is the magnetostatic contribution in a constant applied field. Employing Eq.~\eqref{eq:Appendix_magnetic_field} in Eq.~\eqref{eq:Appendix_free_energy_Average}, the average Gibbs free energy density is obtained as:
\begin{align}\label{eq:Appendix_free_energy_Su}
\frac{G_1}{V_S} & =  \frac{G_0}{V_S} + \frac{\phi_0H_e}{ad} \frac{\mathrm{cosh}\left(\frac{x_v}{\lambda}\right)}{\mathrm{cosh}\left(\frac{d}{2\lambda}\right)} -  \frac{1}{ad}\frac{\phi_0^2}{4\pi \mu_0\lambda^2} \left[ I_1(x_v) - I_2(x_v) \right] , 
\end{align}
where
\begin{align}
\frac{G_0}{V_S} & \equiv \frac{\varepsilon}{ad} - \frac{\lambda\mu_0H_e^2}{d}\mathrm{tanh}\left( \frac{d}{2\lambda} \right) + \frac{\lambda\mu_0H_I^2}{d}\mathrm{coth}\left( \frac{d}{2\lambda} \right) - \frac{\phi_0H_e}{ad}, \label{eq:g0}\\
I_1(x_v) & \equiv  \int_{0}^{\infty} dk \frac{1}{\sqrt{k^2 + \frac{1}{\lambda^2}}}  \left[ 1-\frac{2\mathrm{sinh}\left(  \left\{ \frac{d}{2}-x_v \right\} \sqrt{k^2 + \frac{1}{\lambda^2}} \right) \mathrm{sinh} \left( \left\{ \frac{d}{2}+x_v \right\} \sqrt{k^2+\frac{1}{\lambda^2}} \right)}{\mathrm{sinh}\left( d\sqrt{k^2+\frac{1}{\lambda^2}}\right)} \right], \label{eq:i1} \\
I_2(x_v) & \equiv \sum_{n\neq 0} \int_{-\infty}^{\infty} dk \frac{e^{ikna}}{\sqrt{k^2+\frac{1}{\lambda^2}}} \left[ \frac{\mathrm{sinh} \left( \left\{\frac{d}{2}-x_v \right\} \sqrt{k^2 + \frac{1}{\lambda^2}}  \right) \mathrm{sinh} \left( \left\{ \frac{d}{2} + x_v \right\} \sqrt{k^2 + \frac{1}{\lambda^2}} \right)}{\mathrm{sinh} \left( d \sqrt{k^2+\frac{1}{\lambda^2}} \right)} \right]. \label{eq:i2}
\end{align}
In simplifying and evaluating the integrals in Eq.~\eqref{eq:Appendix_free_energy_Average} using Eq.~\eqref{eq:Appendix_magnetic_field}, we performed integration by parts on the curl terms and replaced $\nabla^2 \pmb{B}$ via Eq.~\eqref{eq:Appendix_magnetic_field_equation}.

The term containing $I_2(x_v)$ [Eq.~\eqref{eq:i2}] represents the energy due to interaction between the different vortices in the chain. Carrying out the integrating using a complex contour and in the limit $d \ll \lambda$, $a \gtrsim d$, we obtain~\cite{Shmidt1970}:
\begin{align}
I_2(x_v) \approx 4 e^{- \frac{\pi a}{d} } \cos^2 \left(\frac{\pi x_v}{d} \right),
\end{align} 
which becomes negligible due to its exponential suppression for $a \gtrsim d$. At low fields relevant to our analysis, $a$ is expected to be much larger than $d$. Thus, we may safely ignore this $I_2(x_v)$ contribution. The $I_{1}(x_v)$ term [Eq.~\eqref{eq:i1}] captures the interaction of the vortices with the surfaces, as explained in the main text. Carrying out the integration after a series expansion in terms of exponential functions~\cite{Shmidt1970}, Eq.~\eqref{eq:i1} is evaluated as:
\begin{align}\label{eq:i1k}
I_{1}(x_v) & = \sum_{n=0}^{\infty} \left[ K_0\left( \frac{2dn + d - 2 x_v }{\lambda} \right) + K_0 \left( \frac{2dn + d + 2 x_v }{\lambda}\right) - 2 K_0\left( \frac{2dn + 2d}{\lambda} \right) \right],
\end{align}
where $K_m(x)$ is the modified Bessel function of the second type and $m$th order~\cite{Abramowitz1968}. 

As we work in the London limit assuming vortices to be point objects, which further implies $\xi \to 0$, our theoretical description is troublesome when the vortices with size $\sim \xi$ come too close to the surfaces, i.e., when $|x_v \pm d/2| \lesssim \xi$. Shmidt~\cite{Shmidt1970} addresses this issue via some approximations and an exact cancellation of the vortex core energy, which is also divergent in the London limit of $\xi \to 0$. We do not need to worry about those details here, except for recognizing that this inadequacy causes $I_{1}(x_v)$ to unphysically diverge around $x_v = \pm d/2$. We resolve this issue by simply redefining $I_1(x_v)$ as:
\begin{align}\label{eq:i1k2}
I_{1}(x_v)  =  & \sum_{n=0}^{\infty}  \left[ K_0\left( \frac{2\xi + 2dn + d - 2 x_v }{\lambda} \right) + K_0 \left( \frac{2\xi + 2dn + d + 2 x_v }{\lambda}\right)  \right. \nonumber \\
 & \left. - 2 K_0\left( \frac{2\xi + 2dn + 2d}{\lambda} \right) \right],
\end{align}
noting that in our London limit $\xi \ll \lambda$, the above expression is practically the same as Eq.~\eqref{eq:i1k} while avoiding unphysical divergences. 

We can express the average Gibbs free energy density [Eq.~\eqref{eq:Appendix_free_energy_Su}] in a normalized form:
\begin{align}\label{eq:gtilde}
\tilde{G}(x_v) & =  \frac{G_{1}}{V_S} \frac{4 \pi a d \mu_0 \lambda^2}{\phi_0^2}  =  \mathrm{const.} + \frac{H_e \ln\left( \lambda/\xi \right)}{H_{c1}} \ \frac{\mathrm{cosh}\left(\frac{x_v}{\lambda}\right)}{\mathrm{cosh}\left(\frac{d}{2\lambda}\right)} - I_{1}(x_v),
\end{align} 
where $H_{c1} = \phi_0 \ln\left(\lambda /\xi \right)/ (4 \pi \mu_0 \lambda^2)$ is the first critical field in the bulk superconductor~\cite{deGennes1966}, we have disregarded the $I_2(x_v)$ term as discussed above, and $I_{1}(x_v)$ is given by Eq.~\eqref{eq:i1k2}. Equation \eqref{eq:gtilde} has been employed in plotting Fig.~\ref{fig:SwoSOC}~(b), while the constant offset has been adjusted to make $\tilde{G}(x_v = \pm d/2) = 0$. 

We now detail the approximations employed in obtaining the expressions for $H^\prime$ [Eq.~\eqref{eq:hprime}], $H_s$ [Eq.~\eqref{eq:hs}], and $H_{I_c}$ [Eq.~\eqref{eq:hic1}] in the main text. As discussed above, we disregard the $I_2(x_v)$ contribution to the Gibbs free energy density [Eq.~\eqref{eq:Appendix_free_energy_Su}] in these evaluations. Furthermore, we consider only the $n=0$ term in $I_{1}(x_v)$ [Eq.~\eqref{eq:i1k2}] as $|K_0(x)|$ decreases with increasing $x$ making the $n>0$ terms smaller. This approximation is very good close to the surface where one of the $K_0(x)$ terms becomes large, but not so good in the middle. Nevertheless, our evaluated $H^\prime$ which is determined by a condition in the middle of the film [Eq.~\eqref{eq:hprime}] differs from Shmidt's result~\cite{Shmidt1970} by a factor close to 1, and thus is a good estimate. In evaluating $H_s$ [Eq.~\eqref{eq:hs}], and $H_{I_c}$ [Eq.~\eqref{eq:hic1}], we work close to the surface such that one of the $n=0$ terms dominates $I_{1}(x_v)$. Our approximation is thus better here and we employ the following properties of the modified Bessel functions of the second type~\cite{Abramowitz1968}:
\begin{align}
\frac{d K_0(x)}{dx} & = - K_1(x), \\
 \frac{d K_1(x)}{dx}  & = \frac{K_1(x)}{x} - K_2(x), \\
\lim_{x \to 0} K_1(x) & = \frac{1}{x}, \\
 \lim_{x \to 0} K_2(x) & = \frac{2}{x^2},
\end{align}  
in evaluating the expressions [Eqs.~\eqref{eq:hs} and \eqref{eq:hic1}] reported in the main text. These are identical to what Shmidt has obtained using a different method~\cite{Shmidt1970}.

\section{Superconductor film with SOC and exchange field}\label{sec:app2}

In this section, we detail the evaluation of average Gibbs free energy density when $\pmb{\alpha} \neq \pmb{0}$. The model and methodology have already been described in Sec.~\ref{sec:SwSOCZ} and we only note certain additional details here. Invoking linearity of the governing equation \eqref{eq:bmain}, the contribution of our assumed $\pmb{\alpha}$ [Eq.~\eqref{eq:alpha_definition}] to the magnetic flux density is obtained as the solution of:
\begin{align}
\lambda^2\nabla^2\pmb{B}-\pmb{B} = &  \left[ \frac{1}{2e} \left\{ \alpha_L\mathrm{cos}\theta_L\delta(x+x_{so}) + \alpha_R\mathrm{cos}\theta_R\delta(x-x_{so}) \right\} \right] \pmb{\hat{z}} \nonumber \\
& -  \left[ \frac{1}{2e}  \left\{ \alpha_L\mathrm{sin}\theta_L\delta(x+x_{so}) + \alpha_R\mathrm{sin}\theta_R\delta(x-x_{so}) \right\} \right] \pmb{\hat{y}},
\end{align}
with the boundary conditions $\pmb{B}(x = \pm d/2) = \pmb{0}$. The result $\pmb{B}_{\alpha}(x)$ has been reported in Eq.~\eqref{eq:balpha}. Employing the total magnetic flux density [Eqs.~\eqref{eq:bcalc1} and \eqref{eq:balpha}] in Eq.~\eqref{eq:gibbs1}, we obtain the total average Gibbs free energy density as described in Eq.~\eqref{eq:gibbswSOC} with
\begin{align}
\frac{G_{\alpha 0}}{V_S} = & - \frac{l_{so}}{d}\frac{(\alpha_L^2 + \alpha_R^2)}{8\mu_0e^2\lambda^2} - \frac{1}{4\mu_0 ed} \left[ \alpha_L \mathrm{cos}\theta_L B_{\alpha,z}(-x_{so}) + \alpha_R \mathrm{cos}\theta_R B_{\alpha,z}(x_{so}) \right] \nonumber \\
& + \frac{1}{4\mu_0 ed} \left[ \alpha_L \mathrm{sin}\theta_L B_{\alpha,y}(-x_{so}) + \alpha_R \mathrm{sin}\theta_R B_{\alpha,y}(x_{so})\right]  \nonumber \\
& + \frac{H_e}{2ed} \left(1-\frac{\mathrm{cosh}(\frac{x_{so}}{\lambda})}{\mathrm{cosh}(\frac{d}{2\lambda})}\right) (\alpha_L\mathrm{cos}\theta_L + \alpha_R\mathrm{cos}\theta_R), \\
\frac{G_{\alpha}}{V_S} = & - \frac{1}{4\mu_0eda}\left[ \alpha_L \mathrm{cos}\theta_L B_{k = 0}(-x_{so}) + \alpha_R \mathrm{cos}\theta_R B_{k = 0}(x_{so})\right] + \frac{\phi_0}{2 \mu_0 a d} B_{\alpha,z}(x_v), \\
  = &  \frac{\phi_0}{\mu_0 a d} B_{\alpha,z}(x_v),
\end{align}
where $B_{\alpha,z}$ and $B_{\alpha,y}$ are the $z$ and $y$ components respectively of $\pmb{B}_{\alpha}$ [Eq.~\eqref{eq:balpha}], and $B_{k}$ is given in Eq. \eqref{eq:Appendix_B_m_and_B_k}.

\bibliography{suFET.bib}

\end{document}